\documentclass[aps,10pt,prd,letterpaper, amsmath, amssymb, preprintnumbers, showpacs, floatfix, nofootinbib, superscriptaddress,tightenlines,twocolumn]{revtex4-1}
\usepackage{tikz}
\usepackage{mystyle_emacs}
\usepackage{stackrel}
\usepackage{theoremref}
\usepackage{placeins}
\usepackage{multirow}
\usepackage{graphicx}
\usepackage{xcolor}

\usetikzlibrary{decorations.markings,decorations.pathmorphing}

\newcommand\mtiny[1]{\mbox{\tiny\ensuremath{#1}}}

\begin{document}

\title{Aspects of Propagator Sparsening in Lattice QCD}

\author{Sam Christian}
\email{samdc@mit.edu}
\affiliation{Center for Theoretical Physics,
Massachusetts Institute of Technology, Cambridge, MA 02139, USA}
\author{William Detmold}
\email{wdetmold@mit.edu}
\affiliation{Center for Theoretical Physics,
Massachusetts Institute of Technology, Cambridge, MA 02139, USA}
%\affiliation{The NSF Institute for Artificial Intelligence and Fundamental Interactions}

\begin{abstract}
In lattice field theory, field sparsening aims to replace quantum fields, or objects constructed from them, with approximations that preserve the appropriate symmetries and maintain many aspects of the physics that the fields determine. For example, an effective sparsening of a quark propagator provides an efficient map from a quark propagator on a fine lattice geometry to a quark propagator defined on a coarser geometry in order to reduce storage and computational costs of subsequent calculational stages while maintaining long-distance correlations and corresponding low-energy physical information. Previous studies have focused on decimating lattice sites or randomly sampling lattice sites to reduce the size of the propagator and subsequent costs of Wick contractions. Here, we extend the study of sparsening to incorporate covariant averaging of spatial sites and examine the effects on two-point and three-point correlation functions involving various hadrons. We find that sparsening is most effective in reproducing the unsparsened versions of these correlation functions when weighted covariant-averaging is sequentially applied many times.
\end{abstract}

\preprint{{MIT-CTP/5826}}
\maketitle

\newpage
\section{Introduction}
\label{sec:intro}
As the lattice spacings used in lattice Quantum Chromodynamics (LQCD) calculations have become increasingly fine, the effects of discretization have been parametrically reduced. However, correlations between nearby lattice sites have become increasingly significant, leading to effectively-redundant computation and storage at many stages of LQCD calculations. In principle, making use of improved actions \cite{Weisz:1982zw,Weisz:1983bn,Symanzik:1983dc,Symanzik:1983gh,Iwasaki:1985we,Iwasaki:1983iya,Luscher:1985zq,Sheikholeslami:1985ij} or perfect actions \cite{Hasenfratz:1993sp,Bietenholz:1995cy,DeGrand:1995ji,Hasenfratz:1998bb} can mitigate this problem as one expects less correlation between neighboring lattice sites at a given level of discretization effects. However, such actions are costly and technically challenging to implement.

A pragmatic way to mitigate this inherent redundancy is to sparsen the spacetime lattice geometry to a coarser grid at some stage in the calculation. Effective sparsening reduces the redundancy by suppressing high-frequency fluctuations while preserving long-distance correlations  that govern low-energy physics. Consider a gauge field $U_\mu(x)$ defined on a lattice geometry $\Lambda$ of size $N_1\times\ldots\times N_d$. Sparsening at this stage consists of defining a map $\mathcal{G}(U_\mu)\mapsto U_\mu'$ that maps the gauge field to a gauge field $U_\mu'$ defined on a second lattice geometry $\Lambda'$ of smaller size $N_1'\times\ldots \times N_d'$. If $\mathcal{G}$ is an effective sparsening, quantities computed from an ensemble of $U_\mu'$ degrees of freedom will preserve low energy physics (see Refs.~\cite{Endres:2015yca,Detmold:2016rnh,Detmold:2018zgk,Shanahan:2018vcv,Abbott:2024knk} for some exploratory works that investigate this direction). It is also possible to apply sparsening to functions of the underlying gauge field such as the quark propagator $S(x|y)$ computed as the inverse of the Dirac operator on a background gauge field. In this case, the propagator computed on the full lattice geometry $\Lambda$ is mapped to a propagator $S'(x|y)$ defined on the coarse geometry $\Lambda'$ via a map $\mathcal{F}(S,U_\mu)\mapsto S'(x|y)$.

In this study, we focus on sparsening of the propagator through the construction of the map $\mathcal{F}$, as has been previously studied in Refs.~\cite{Detmold:2019fbk,Li:2020hbj}. Ref.~\cite{Detmold:2019fbk} presented studies of  sparsening via decimation, where the propagator on the reduced lattice geometry was obtained by decimation: $S'(x|y)=S(x|y)\big|_{y\in \Lambda'}$, with $\Lambda'< \Lambda$ constant across all gauge configurations. The sparsened propagators were subsequently used to study hadronic and nuclear correlation functions. Li \textit{et al.} \cite{Li:2020hbj} studied sparsening in the context of random field selections, where $S'(x|y)=S(x|y)\big|_{x,y \in \Lambda'(U)}$ and $\Lambda'(U)$ varies randomly between gauge configurations. They applied this method to both two-point and three-point correlation functions for mesons.

Since the computational cost of Wick contractions increases factorially with the number of quark fields required in interpolating operators, perhaps the clearest use case for propagator sparsening is to reduce the cost of correlation functions of multiple nucleons \cite{amarasinghe2023variational,Detmold:2024iwz} and other many-body systems \cite{Abbott:2023coj,Abbott:2024vhj}. For example, Ref. \cite{amarasinghe2023variational} used sparsening to reduce the computationally-intensive momentum projection in the correlation functions of dibaryon operators, whose computational cost scales as $\mathcal{O}(V^3)$ with spatial volume $V$. Additionally, Ref. \cite{feng2022lattice} studied the electromagnetic contribution to the mass splitting between $\pi^+$ and $\pi^0$ mesons, which involves a sum over $\mathcal{O}(V^2)$ contributions, using the random sparsening techniques developed in Ref.~\cite{Li:2020hbj} to reduce the computational cost.

In this work, we study an extended form of propagator sparsening that averages neighboring sites gauge-covariantly. Throughout, we refer to the projection of the propagator to a reduced spatial geometry as \textit{decimation} and gauge-covariantly averaging as \textit{blocking}, while the overall process is referred to as \textit{sparsening}. This more elaborate sparsening may be more effective at preserving long-distance correlations in derived correlation functions and, given the averaging can be performed in many ways, provides significant opportunities for optimization.

We begin by introducing general aspects of sparsening and the particular details of the approach we use in this work (Sec. \ref{sec:sparse}). We then discuss the setup for our experiments in Sec.~\ref{sec:latticeDetails} and investigate the effects of sparsening on two-point correlation functions and resulting ground state energies (Sec. \ref{sec:twopoint}). In Sec. \ref{sec:three}, we expand this analysis to three-point correlation functions involving the vector current and the resulting form factors at varying momenta.\footnote{The computations in this work are performed using the gpt \cite{gpt} python package}

\section{Propagator sparsening}
\label{sec:sparse}

We begin by defining the concept of quark-propagator sparsening in some generality before specializing to the approach used in this work. Sparsening of other field objects can be considered similarly.
Conceptually, sparsening implies a localized reduction of the information contained in a field or product thereof, with decimation \cite{Detmold:2019fbk} being the simplest example. 
Consider a $d$-dimensional lattice geometry $\Lambda=\{a=(a_1,\ldots a_d);a_i\in \mathbb{Z},0\leq a_i < N_i\}$ where $N_i$ is the  extent in the $i$th dimension. Consider also a decimated lattice geometry $\Lambda'=\{a=({a}_1s_1,\ldots a_ds_d);{a}_i\in\mathbb{Z},0\leq {a}_i < N_i/s_i\}$, where $s_i$ is the sparsening factor in the $i$th dimension (a divisor of $N_i$). 
There are many possible sparsening constructions and it is conceptually useful to think of them as two-step processes.
First, a blocking step $\mathcal{F}(S,U_\mu)\mapsto \widetilde{S}$ is applied to the propagator $S$ and gauge field $U_\mu$ defined on lattice geometry $\Lambda$ such that $\widetilde{S}$ is also defined on $\Lambda$. Subsequently, the propagator is projected to the decimated geometry $\Lambda'$: $ S'(x|y)=\widetilde{S}(x|y)\big|_{x,y\in \Lambda'}$. 

In the applications we investigate below, we will consider only isotropic spatial sparsening, in which $s_i=s$ for all spatial dimensions and $s_d=1$, and a spatially isotropic geometry with $N_i=N$ for $i\in\{1,\ldots,d-1\}$ and $N_d=N_t$.

\subsection{Decimation}
\label{sec:idealLimit}

In applications to momentum-space correlation functions that involve quark propagators, a decimation amounts to a partial momentum-projection of the operators in the correlation function, in which  higher momentum-modes are not projected out in the Fourier transform to momentum space. Therefore, as we will see explicitly below, decimation typically induces additional excited-state contamination. However, when the large separation limit of the correlation function is taken, the larger-momentum states typically decay rapidly because of their higher energy, leaving only the lowest-momentum state.

As an explicit example, consider a pion two-point correlation function in the limit of degenerate quark masses $\langle \pi^\dagger(\vec{x},0)\pi(\vec{0},0)\rangle= \langle S^\dagger(\vec{x}|0) S(\vec{x}|0)\rangle $, where $\pi(x)=\overline{d}(x)\gamma_5 u(x)$ is an interpolating operator that has nonzero overlap with the lowest-energy state with the quantum numbers of the pion. In this discussion, we denote the set of spatial components of each vector in $\Lambda$ as $\Lambda_3\equiv \{\vec{a}=(a_1,a_2,a_3);{a}_i\in \mathbb{Z},0\leq {a}_i<N\}$.

In order to construct the correlation function of the momentum-space pion operator, $\pi(\vec{p},t)=\sum_{\vec{x}\in \Lambda_3}e^{-i\vec{x}\cdot \vec{p}}\pi(\vec{x},t) $, we take the discrete Fourier transform of the position-space correlation function:
\begin{align}
    &\left\langle \left(\sum_{\vec{x}\in \Lambda_3}e^{-i\vec{x}\cdot \vec{p}}\pi^\dagger (\vec{x},t)\right)\pi(\vec{x}=\vec{0},0)\right\rangle\\
    &\hspace*{2cm}\propto\langle \pi^\dagger(\vec{p},t)\sum_{\vec{p'}\in\tilde{\Lambda}_3}\pi(\vec{p}',0)\rangle\notag\\
    &\hspace*{2cm}=\langle \pi^\dagger (\vec{p},t)\pi(\vec{p},0)\rangle\rangle, \notag
\end{align}
where we have used the relation that a correlation function between operators of different three-momenta at different times is zero and $\tilde{\Lambda}_3=\{\frac{2\pi \vec{a}}{N};a_i\in \mathbb{Z},0\leq a_i<N\}$ is the conjugate lattice of three-momenta. 

When applying decimation, we project $\pi(\vec{x},t)$ using a three-dimensional sparse grid $\Lambda'_3$ as $\sum_{\vec{x}\in \Lambda'_3} e^{-i\vec{x}\cdot \vec{p}}\pi(\vec{x},t) $. For a given momentum $\vec{p}$, we denote the corresponding vector of integers $\vec{a}_{\vec{p}}\equiv \frac{N}{2\pi}\vec{p}$. The correlation function in momentum space then becomes
\begin{align}
\label{eq:theoreticalExpectation}
    \langle\pi^\dagger (\vec{p},t)\pi(\vec{p},0)\rangle_{\Lambda'}&= \sum_{\vec{x}\in \Lambda_3'}\sum_{\vec{p}'\in\tilde{\Lambda}_3}e^{-i\vec{x}\cdot(\vec{p}-\vec{p}')}\langle \pi^\dagger (\vec{p}',t)\pi(\vec{p}',0)\rangle\\
   &\hspace*{-1cm}=\sum_{\vec{x}\in\Lambda_3'}\sum_{\vec{p}'\in \tilde{\Lambda}_3}e^{-2\pi i \frac{\vec{x}\cdot (\vec{a}_{\vec{p}}-\vec{a}_{\vec{p}'})}{N}}\langle \pi^\dagger (\vec{p}',t)\pi(\vec{p}',0)\rangle\notag \\
  % &=\sum_{p'\in \tilde{\Lambda},p'\neq p}\prod_j\bigg[\left(\frac{1-e^{-i\frac{2\pi n_j}{N}(a_j-a'_j)}}{1-e^{-i(p_j-p'_j)}}\right)\\
  &\hspace*{-1cm}=\sum_{\vec{l}\in A_3} \langle \pi^\dagger (\vec{l},t)\pi(\vec{l},0)\rangle\notag
  \label{eq:secondStep}
\end{align}
for
\begin{align}
   A_3&\equiv \left\{\frac{2\pi}{N}[(a_{\vec{p}})_i+c_is];c_i\in \mathbb{Z},0\leq (a_{\vec{p}})_i+c_is<N\right\},
\end{align}
where we have added a subscript $\Lambda'$ to the correlation function to denote that only a partial discrete Fourier transformation has been performed. As $t\to \infty$, $\langle \pi^\dagger (l,t)\pi(l,0)\rangle\to e^{-E'(\vec{l})t}$, and the lowest-energy state with the allowed quantum numbers, corresponding to $E'(\vec{l})$, will dominate. 
However, the lowest-energy state on the decimated lattice is not always the lowest-energy state of the undecimated lattice. Denote the dispersion relation on the undecimated lattice as $E(\vec{l})$. The undecimated dispersion relation as a function of a given Euclidean momentum-component ${l}_i$ is symmetric around components of momentum ${l}_i=\pi$: that is, $E({l}_i,l_\perp)=E(2\pi-\vec{l}_i,l_\perp)$, where ``$l_\perp$" represents the dependence of $E$ on the components of momentum orthogonal to the $i$th Cartesian-direction.
Because of the decimation, the effective dispersion relation on the decimated lattice is
\begin{align}
    E'({l}_i,l_\perp)&=\begin{cases}
        E\left({l}_i\!\!\!\!\mod \frac{2\pi}{s},l_\perp\right), &{l}_i\!\!\!\! \mod \frac{2\pi}{s}\leq \frac{\pi}{s}\\
        E\left(\frac{2\pi}{s}-\left[{l}_i\!\! \!\!\mod \frac{2\pi}{s}\right],l_\perp\right),&{l}_i\!\!\!\!\mod \frac{2\pi}{s}\geq \frac{\pi}{s}.
    \end{cases}
\end{align}
From the dispersion relation on the decimated lattice, we see that the decimated correlation functions are therefore asymptotically sensitive to energies of states with any momenta $\vec{p}$ such that each component of $\vec{p}$ is less than or equal to $\frac{\pi}{s}$.

This analysis holds for any local choice of interpolating operator and also for higher-point correlation functions. However, more care must be taken in analyzing quantities, such as form factors, that depend not just on the energies extracted from correlation functions but also on their overall normalization. This normalization can be altered by decimation and is discussed in Sec.~\ref{sec:threePointSetup}.

\subsection{Nearest and next-to-nearest-neighbor blocking}
\label{sec:blockingDescription}

In this work, we study an extension of decimation by gauge-covariantly averaging -- or blocking -- the propagator before decimating.
Specifically, we consider blocking that involves nearest-neighbor and next-to-nearest-neighbor lattice sites of the form,
\begin{widetext}\begin{equation}\begin{aligned}
    \mathcal{F}[S(x|y)]=&S(x|y)+\frac{\alpha}{6}\sum_{\mu_1}U_{\mu_1}(x)S(x+\mu_1|y)
    +\frac{\alpha'}{6}\sum_{\mu_2} S(x|y+\mu_2)U_{\mu_2}^\dagger(y)%\\
    +\frac{\alpha\alpha'}{36}\sum_{\mu_1,\mu_2}U_{\mu_1}(x)S(x+\mu_1|y+\mu_2)U_{\mu_2}^\dagger(y)\\
    &+\frac{\beta}{24}\sum_{\mtiny{\begin{array}{c}\mu_1\neq\mu_1'\\ \mu_1\neq-\mu_1'\end{array}}}U_{\mu_1}(x)U_{\mu_1'}(x+\mu_1)S(x+\mu_1+\mu_1'|y)
    +\frac{\beta'}{24}\sum_{\mtiny{\begin{array}{c}\mu_2\neq\mu_2'\\ \mu_2\neq-\mu_2'\end{array}}}S(x|y+\mu_2+\mu_2')U_{\mu_2}^\dagger(y)U_{\mu_2'}^\dagger(y+\mu_2)\\
    &+\frac{\beta\beta'}{576}\sum_{\mtiny{\begin{array}{c}\mu_1\neq\mu_1', \mu_1\neq-\mu_1'\\\mu_2\neq\mu_2', \mu_2\neq-\mu_2'\end{array}}}U_{\mu_1}(x)U_{\mu_1'}(x+\mu_1)
    S(x+\mu_1+\mu_1'|y+\mu_2+\mu_2')U_{\mu_2}^\dagger(y)U_{\mu_2'}^\dagger(y+\mu_2)\\
    &+\frac{\beta\alpha'}{144}\sum_{\mu_1\neq \mu_1',\mu_2}U_{\mu_1}(x)U_{\mu_1'}(x+\mu_1)
    S(x+\mu_1+\mu_1'|y+\mu_2)U^\dagger_{\mu_2}(x)\\
    &+\frac{\alpha\beta'}{144}\sum_{\mu_1,\mu_2\neq\mu_2'}U_{\mu_1}(x)S(x+\mu_1|y+\mu_2+\mu_2')U^\dagger_{\mu_2}(x)U_{\mu_2'}^\dagger(y+\mu_2),
\end{aligned}\label{eq:propagatorSparsening}\end{equation}
\end{widetext}
 with $\mu_1,\mu_1',\mu_2,\mu_2'\in\{-3,-2,-1,1,2,3\}$ and with $U_{-\mu}\equiv U_\mu^\dagger (x-\hat{\mu}),\mu>0$.
The normalization of the  coefficients of the nearest-neighbor and next-to-nearest-neighbor terms, $\alpha^{(\prime)},\ \beta^{(\prime)}$, are chosen so that if $\alpha^{(\prime)}=1, \beta^{(\prime)}=1$, the 6 points that are nearest neighbors to $x$ (to $y$) are given a total weight equal to that of $x$ (of $y$), and similarly for the 24 next-to-nearest-neighbor points. When sparsening symmetrically at both the source and sink, we take $\alpha'=\alpha, \beta'=\beta$.
In order that all next-to-nearest neighbors are the same distance from the central point, we exclude $\mu_i=\{\mu_i',-\mu_i'\}$ from the sums, as indicated.

We also consider multiple sequential applications of Eq. (\ref{eq:propagatorSparsening}) as
\begin{align}
\label{eq:multipleBlockingSteps}
    \mathcal{F}^n[S(x|y)]&=\underbrace{\mathcal{F}\circ...\circ\mathcal{F}}_{\times n}[S(x|y)],
\end{align}
where $n$ is the number of blocking steps. Iterative Wuppertal smearing \cite{Daniel:1992ek}, which applies a discretization of the Klein-Gordon operator to quark fields, corresponds to this smearing procedure with $\alpha=1,\beta=0$.

As $n$ is increased, all paths of gauge links between any two points $x$ and  $x'$ are eventually included in $\mathcal{F}^n$. If the length of the path is $l$ links and $\beta=0$, then the weight of the path in its contribution to $\mathcal{F}^n$ is $\left(\frac{\alpha}{6}\right)^l$. If in a given path the spatial direction of neighboring links is different, then that path has a joint between those neighboring links. Let $j$ denote the total number of joints in a given path. Then the weight of any single path for nonzero $\beta$ is
$\sum_{i=0}^{j}\binom{j}{i}\left(\frac{\alpha}{6}\right)^{l-2i}\left(\frac{\beta}{24}\right)^{i}$.
Between paths of equal length but with different numbers of joints, introducing nonzero positive $\beta$ has the effect of giving more weight to the path with a greater number of joints.
If we block the propagator sink in the manner of Eq. (\ref{eq:propagatorSparsening}), and then decimate $S(x|y)$ by a factor of $s\geq 3$, then 19 out of every $s^3$ lattice sites contributes some weight to $S'(x|y)$. More relevant information is potentially preserved in this procedure than in decimation without blocking.

\subsection{Correlation function construction}
\label{sec:corr_const}
As an example of the effect of this blocking, we consider again the case of the pion two-point correlation function. 
Instead of using the local pion interpolating operator $\overline{d}(x)\gamma_5 u(x)$, we consider the operator
\begin{align}
\label{eq:modifiedPionOperator}
    \pi'(x;\alpha)=\overline{d}(x)\gamma_5 u(x)+\alpha\sum_\mu \overline{d}(x)U_\mu(x)u(x+\hat{\mu})\\
    +\alpha\sum_\mu \overline{d}(x+\hat{\mu})U^\dagger_\mu(x)u(x)\notag\\
    +\alpha^2\sum_{\mu,\mu'}\overline{d}(x+\hat{\mu}')U^\dagger_{\mu'}(x)U_\mu(x)u(x+\hat{\mu}),\notag
\end{align}
with $\mu,\mu' \in \{1,2,3,-1,-2,-3\}$. Projecting to a given three-momentum gives
$\pi'(\vec{p},t;\alpha)=\sum_{\vec{x}} e^{-i\vec{p}\cdot\vec{x}}\pi'(x;\alpha)$.

The two-point correlation function can then be computed as
\begin{align}
    \label{eq:modifiedPionCorrelationFunction}
    &\langle\pi'(\vec{p},t;\alpha)^\dagger \pi'(\vec{0};\alpha')\rangle=\sum_{\vec{x}\in\Lambda}e^{i\vec{p}\cdot \vec{x}}\Tr\bigg[\bigg(S^d(0|\vec{x},t)\\
    &+\alpha\sum_{\mu_1'}S^d(0|\vec{x}+\hat{\mu}_1',t)U_{\mu_1'}^\dagger(\vec{x},t)\notag\\
    &+\alpha'\sum_{\mu_2'}U_{\mu_2'}(0)S^d(\vec{0}+\hat{\mu}'_2,0|\vec{x},t)\notag\\
    &+\alpha\alpha'\sum_{\mu_1'\mu_2'}U_{\mu_2'}(0)S^d(\vec{0}+\hat{\mu}_2',0|\vec{x}+\hat{\mu}_1',t)U_{\mu_1'}^\dagger(\vec{x},t))\bigg)\gamma_5\notag\\
    &\times\bigg(S^u(\vec{x},t|0)+\alpha\sum_{\mu_1} U_{\mu_1}(\vec{x},t)S^u(\vec{x}+\hat{\mu}_1,t|0)\notag\\
    &+\alpha'\sum_{\mu_2}S^u(\vec{x},t|0+\hat{\mu}_2)U_{\mu_2}^\dagger(0)\notag\\
    &+\alpha\alpha'\sum_{\mu_1,\mu_2}U_{\mu_1}(\vec{x},t)S^u(\vec{x}+\hat{\mu}_1,t|0+\hat{\mu}_2)U_{\mu_2}^\dagger(0)\bigg)\gamma_5\bigg].\notag
\end{align}

The form of Eq.~\eqref{eq:modifiedPionCorrelationFunction} is exactly the result of gauge-covariant averaging with $\beta=\beta^\prime=0$ in Eq.~\eqref{eq:propagatorSparsening}.
\begin{figure}[!ht]
    \centering
\includegraphics[width=0.7\columnwidth]{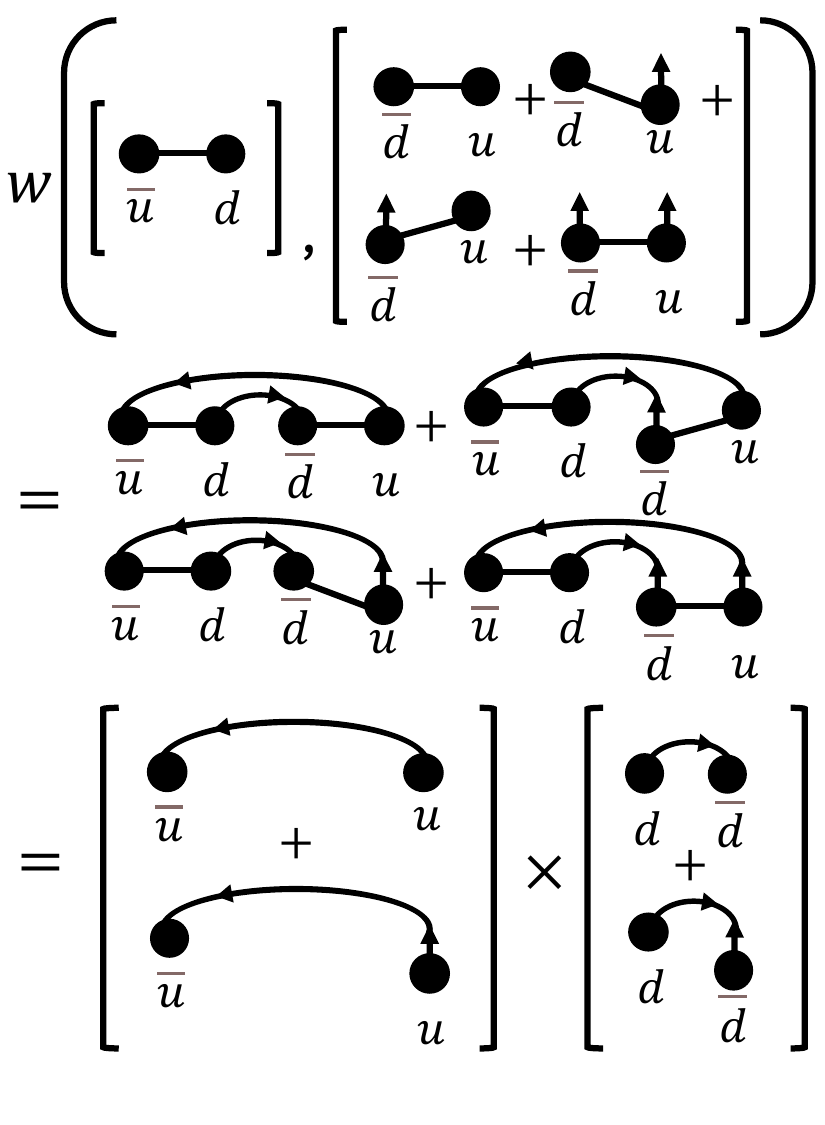}
    \caption{A diagrammatic demonstration of the relation between Eqs. (\ref{eq:propagatorSparsening}) and (\ref{eq:modifiedPionCorrelationFunction}). $\overline{d}\mbox{---}u$ represents the pion operator $\pi(x)$, while $\overline{d}\mbox{---}u\uparrow$ (the upper right object with $u$ shifted down and a vertical arrow emanating from $u$) represents the non-local operator $\overline{d}(x)U_\mu(x)u(x+\mu)$. $\overline{u}\leftarrow u$ represents $S^u(0|x)$, while $\overline{u}\leftarrow \uparrow u$  (the lower left object with a line with an arrow connecting $\overline{u},u$ and a vertical arrow emanating from $u$) represents $S^u(0|x)U_\mu(x)$.
    The symbol $w$ represents Wick contraction of quark operators, while the multiplication symbol $\times$ represents matrix multiplication of propagators.}
    \label{fig:pionDiagram}
\end{figure}
This equivalence is straightforward to see in a diagrammatic representation; for the case of blocking only the sink ($\alpha'=\beta=\beta'=0$), we show this explicitly in Fig. \ref{fig:pionDiagram}. The equivalence when $\beta\neq 0$ and when source blocking is also applied follows analogously. This diagrammatic representation also suggests why this correspondence is true for any local interpolating operator. Explicitly, if an operator is a function of various local quark fields $\{q_i\}$, blocking the propagator by covariantly-averaging over nearest-neighbors corresponds to modifying the operator as
\begin{align}
\label{eq:nonlocalOperatorModification}
    \mathcal{O}(\{q_i(x)\})\mapsto \mathcal{O}(\{q_i(x)+\alpha\sum_\mu  U_\mu(x) q_i(x+\mu)\}).
\end{align}

\subsection{Sink sparsening}
When computing two-point correlation functions, one would ideally treat the source and sink identically in order to enforce convexity of correlation functions. However, propagator inversion is computationally expensive and below we will study many different sparsening schemes. In order that we can achieve this with available computational resources, we also study sparsening of only the sink. 
When sparsening the sink but not the source, Eq. (\ref{eq:propagatorSparsening}) with $\alpha'=\beta'=0$ reduces to
\begin{multline}
    \mathcal{F}[S(x|y)]=S(x|y)+\frac{\alpha}{6}\sum_{\mu}U_{\mu}(x)S(x+\mu|y)\\
    +\frac{\beta}{24}\sum_{\mu\neq\mu', \mu\neq-\mu'}U_\mu(x)U_{\mu'}(x+\mu)S(x+\mu+\mu'|y),
\end{multline}
and correlation functions of the form of Eq.~(\ref{eq:modifiedPionCorrelationFunction}), $\langle \mathcal{O}'_\alpha(\vec{p})\mathcal{O}'_{\alpha'}(0)\rangle$, reduce to $\langle \mathcal{O}'_\alpha(\vec{p})\mathcal{O}(0)\rangle$, where $\mathcal{O}'_\alpha$ is the smeared operator, and $\mathcal{O}$ is the unsmeared operator.

\section{Lattice details}
\label{sec:latticeDetails}
We perform the computations in this study on a $24^3\times 48$ lattice geometry. We use a smaller lattice than in Ref. \cite{Detmold:2019fbk} so that we can perform a larger set of analyses efficiently. Additionally, a spatial lattice size of $L=24$ has more divisors and therefore results can be computed for more decimations. We use a tree-level improved L\"uscher-Weisz  gauge action \cite{Luscher:1985zq} with $\beta=6.1$ and one level of stout smearing \cite{Morningstar:2003gk} with $\rho=0.125$ and a Wilson-clover fermion action \cite{Sheikholeslami:1985ij} with an unrenormalized quark mass of $m_0=-0.245$, $c_{sw}=1.249$, and $N_f=3$ degenerate flavors of quarks. We use $N_{\rm cfg}=100$ gauge configurations.

We compute quark propagators from a single point source on each configuration and decimate them by factors of $s\in\{2, 3, 4, 6, 8, 12\}$. Measurement errors are computed using the bootstrap procedure: we draw 100 gauge configurations with replacement 10,000 times and calculate the quantities of interest each time. Then we use the standard deviation of the resulting distribution as the measurement uncertainty.

\section{Two-point correlation functions}
\label{sec:twopoint}

\subsection{Details of lattice calculations}
\label{sec:twopointSetup}

We compute two-point correlation functions of the pion and proton for momenta $\vec{p}\in \frac{2\pi}{N}\{[0,0,0],[1,0,0],[2,0,0]\}$, as well as for the $\Delta$ baryon for zero momentum. For the pion, we use the interpolating operator
\begin{align}
    \pi(x)=\overline{u}_a (x)\gamma_5 d_a(x),
\end{align}
and for the proton, we use
\begin{align}
    N(x)=\epsilon_{abc}u^T_a(x)(u_b(x) C\gamma_5 d_c(x),
\end{align}
where color indices are shown explicitly and summed over, $\gamma_5=\gamma_1\gamma_2\gamma_3\gamma_4$ is the product of Euclidean Dirac matrices $\gamma_\mu$ and $C=i\gamma_2\gamma_4$ is the charge conjugation operator.
For the $\Delta$ baryon, we use \cite{zanotti2003spin}:
\begin{align}
    \Delta_\mu(x)&= \epsilon_{abc}(u^{T}_a(x)C \gamma_\mu d_b(x))\gamma_5 u_c(x),
\end{align}
where $\mu\in \{1,2,3\}$.

For the $\Delta$ baryon two-point correlation function at zero momentum, we project onto states with definite spin-$\frac{3}{2}$ using the projection operator \cite{Hockley:2023xms}
\begin{align}
    P_{\mu\nu}^{3/2}&=(\eta_{\mu\nu}-\gamma_\mu\gamma_\nu)\delta_{\mu\neq 4,\nu\neq 4}
\end{align}
(this simplified expression only holds for zero momentum).

Explicitly, we then compute the correlation functions as
\begin{align}
    C^{\pi}(t,\vec{p})&=\sum_{\vec{x}}e^{-i\vec{x}\cdot \vec{p}}\,\Tr[\pi^\dagger(\vec{x},t)\pi(\vec{x},t)],\\
    C^{N}(t,\vec{p})&=\sum_{\vec{x}}e^{-i\vec{x}\cdot \vec{p}}\,\Tr[PN^\dagger(\vec{x},t)N(\vec{x},t)],\\
    C^{\Delta}(t,\vec{0})&=\sum_{\vec{x}}\,\Tr[P\Delta_{2}^\dagger(\vec{x},t)\Delta_\lambda(0)P^{3/2}_{\lambda2} ],
\end{align}
where $P=\frac{1}{2}(I+\gamma_0)$ is the parity projection operator and $\Tr$ denotes the Hilbert space and spin trace.

\subsection{Metrics for measuring the effects of modifying interpolating operator}
\label{sec:methodsOfSuccess}

In the discussion below, statistical analysis procedures are applied to both the sparsened and unsparsened quantities. We differentiate between the two when necessary using a $s$ or $us$ subscript, respectively, but otherwise leave this dependence implicit.
We compute the effective energy function for hadron $h$ with momentum $\vec{p}$ at time $t$ using the expression
\begin{align}
E^h(t,\vec{p})=
\begin{cases}
    \frac{\text{arcsinh}[C^h(t-1,\vec{p})]-\text{arcsinh}[C^h(t+1,\vec{p})]}{2C^h(t,\vec{p})}&h\in\{\pi\}\\
    \frac{\text{arccosh}[C^h(t-1,\vec{p})]+\text{arccosh}[C^h(t+1,\vec{p})]}{2C^h(t,\vec{p})}&h\in \{N,\Delta\}.
\end{cases}
\end{align}

In order to find the energy associated to a corresponding momentum, we adopt a simplified version of the fitting procedure of Ref. \cite{NPLQCD:2020ozd}. We denote the effective energy for a given bootstrap sample $i$ as $E_{i}^h(t,\vec{p})$. For a given time-range $[t_{-},t_{+}]$ , for each bootstrap sample $i$ we compute 
\begin{align}
    (\chi^2_{i,t_{-},t_{+}})^{h,\vec{p}}(E)&=\sum_{t=t_-}^{t_+}\frac{(E_i^h(t,\vec{p})-E)^2}{(\sigma^{h,\vec{p}}(t))^2},
\end{align}
where $\sigma^{h,\vec{p}}(t)$ is the bootstrapped standard deviation of the effective energy at time $t$. We take $E^{h,\vec{p}}_{i,\text{min},t_-,t_+}$ to be the value of $E$ that minimizes this $\chi^2$ for a given bootstrap sample $i$. We then compute $E^{h,\vec{p}}_{\text{min},t_{-},t_{+}}$ as the mean value of the $E^{h,\vec{p}}_{i,\text{min},t_{-},t_{+}}$, and $\sigma^{h,\vec{p}}_{\text{min},t_{-},t_{+}}=Q_{5/6}(E^{h,\vec{p}}_{i,\text{min},t_{-},t_{+}})-Q_{1/6}(E^{h,\vec{p}}_{i,\text{min},t_{-},t_{+}})$, where $Q_a$ is the $a$th quantile of the bootstrap distribution.

We then compute a mean $\chi^2$ as
\begin{align}
   (\chi^2_{t_{-},t_{+}})^{h,\vec{p}}&=\sum_{t=t_{-}}^{t_{+}}\frac{(E^h(t,\vec{p})-E^{h,\vec{p}}_{\text{min},t_{-},t_{+}})^2}{(\sigma^{h,\vec{p}}(t))^2}
\end{align}
and reject any time-ranges $[t_{-},t_{+}]$ for which  $(\chi^2_{t_{-},t_{+}})^{h,\vec{p}}>2$. For each time window, we compute the corresponding $p$-value as $p_{t_{-},t_{+}}^{h,\vec{p}}=1-\text{CDF}_{\chi^2}((\chi^2_{t_{-},t_{+}})^{h,\vec{p}})$, where $\text{CDF}_{\chi^2}$ is the cumulative distribution function of the $\chi^2$ distribution with the number of degrees of freedom being the number of time-steps. Finally, we take weighted means of these quantities as a function of the fit-range boundaries, $t_{-}$ and $t_{+}$. The energy plateau  $E^{h,\vec{p}}_p$ and standard deviation $\sigma^{h,\vec{p}}_p$ we report are constructed as
\begin{align}
    w_{t_{-},t_{+}}^{h,\vec{p}}&=\frac{1}{{\cal N}}\frac{p_{t_{-},t_{+}}^{h,\vec{p}}}{(\sigma_{\text{min},t_{-},t_{+}}^{h,\vec{p}})^2},\\
    \label{eq:Ep}
    E_{p}^{h,\vec{p}}&=\sum_{t_{-}\geq 2,t_{+}\geq t_{-}+2}E_{\text{min},t_{-},t_{+}}^{h,\vec{p}} w_{t_{-},t_{+}}^{h,\vec{p}},\\
    (\sigma_{\text{stat}}^{h,\vec{p}} )^2&=\sum_{t_{-}\geq 2,t_{+}\geq t_{-}+2}(\sigma^{h,\vec{p}})^2_{\text{min},t_{-},t_{+}}w^{h,\vec{p}}_{t_{-},t_{+}},\\
    (\sigma_{\text{syst}}^{h,\vec{p}})^2&=\sum_{t_{-}\geq 2,t_{+}\geq t_{-}+2}(E^{h,\vec{p}}_{\text{min},t_{-},t_{+}}-E^{h,\vec{p}}_p)^2 w^{h,\vec{p}}_{t_{-},t_{+}}\notag,\\
   (\sigma_p^{h,\vec{p}})^2&=(\sigma_{\text{stat}}^{h,\vec{p}})^2+(\sigma_{\text{syst}}^{h,\vec{p}})^2,
\end{align}
where the normalization factor ${\cal N}$ is such that the weights sum to one and $\sigma^{h,\vec{p}}_{\text{stat}},\sigma^{h,\vec{p}}_{\text{syst}}$ are the statistical and systematic uncertainties, respectively.

We consider two main approaches to defining the effectiveness of the sparsening procedure. One is to compute the distance of the sparsened $E_s^h(t,\vec{p})$ from the unsparsened $E_{us}^h(t,\vec{p})$. Another is to compute the difference of the sparsened $E_s^h(t,\vec{p})$ from the ground state energy. For the latter metric, one must be careful if only blocking the sink as in that case, the correlation function is no longer convex. Therefore, it is not necessarily best to optimize recovery of the ground state, as there is no way to tell from the sparsened correlator alone that this is the genuine ground-state energy or instead arises from cancellations between different states. However, the second approach can be used to optimize for sparsening procedures that have less excited-state contamination when both the source and  sink are sparsened.
It is also desirable to minimize the uncertainty in the sparsened correlation functions, as compared to the unsparsened correlation function.

Specifically, we focus on three different metrics that measure the efficacy of sparsened quantities in reproducing the unsparsened quantities:

\begin{enumerate}
    \item The difference in effective energy between the unsparsened effective energy and sparsened effective energy at all times less than a fixed maximum time, $\hat t$, weighted by the absolute error of the two:
    \begin{align}
    \label{eq:metricone}
        \mathcal{M}_1^{h,\vec{p}}(\hat t)&=\sum_{t=1}^{\hat t} \frac{(E_{us}^h(t,\vec{p})-E_{s}^h(t,\vec{p}))^2}{(\sigma_{us}^{h,\vec{p}}(t))^2+(\sigma_s^{h,\vec{p}}(t))^2}.
    \end{align}
    \item The difference between extracted energy (from Eq. \eqref{eq:Ep}) and sparsened effective energy, weighted by the absolute error of the two:
    \begin{align}
    \label{eq:metrictwo}
        \mathcal{M}_2^{h,\vec{p}}({\hat t})&=\sum_{t=1}^{\hat t} \frac{(E_{s}^h(t,\vec{p})-E_p^{h,\vec{p}})^2}{(\sigma_s^{h,\vec{p}}(t))^2+(\sigma_p^h)^2}.
    \end{align}
    We also compare $\mathcal{M}_2^h$ to the same quantity computed from the unsparsened correlation function;
    \item The summed ratio of the squared standard deviations of the unsparsened and sparsened correlation functions at all times less than a maximum time.
    \begin{align}
    \label{eq:M3}
        \mathcal{M}_3^{h,\vec{p}}(\hat{t})&=\sum_{t=1}^{\hat{t}} \frac{(\sigma_s^{h,\vec{p}}(t))^{2}}{(\sigma_{us}^{h,\vec{p}}(t))^2}.
    \end{align}
\end{enumerate}

For each correlation function, we define $t'$ as the minimum $t$ for which $\sigma_{us}^{h,\vec{p}}/E_{us}^h(t,\vec{p})>s_{\text{max}}=0.5$ and then use it to set $\hat{t}=\rm{min}(t',17)$.
For a subset of our results, we varied the chosen cutoff $s_{\text{max}}$ and found no significant dependence of our results on its value, provided that $s_{\text{max}}\simeq 1$.

\subsection{Results}
\label{sec:twoPointResults}
We compute these correlation functions and metrics for the pion and proton two-point correlation functions for momenta $\vec{p}\in\frac{2\pi}{N}\{[0,0,0],[1,0,0],[2,0,0]\}$, as well as the $\Delta$ two-point correlation function for zero momentum. Correlation functions are computed for sinks constructed for seven decimation factors (including the undecimated case, $s=1$) and for blocking templates $\mathcal{F}^1$ and $\mathcal{F}^5$ (see Eq.~\eqref{eq:propagatorSparsening}), varying $\alpha,\beta$ over the range $[-1,1]\times [-1,1]$. 
For a single coupling value ($\alpha=\beta=1$), we also compute $\mathcal{F}^{20}$. For $\alpha=\beta=1$, we study symmetric blocking of the source and sink using $\mathcal{F}^1,\mathcal{F}^5,$ and $\mathcal{F}^{20}$. 

The large number of sparsening parameters that can vary in our studies lead to many possible ways to present the data. In the main text, we present examples that highlight trends that are present, with a more extensive set of results available in the Supplementary Material.

\begin{figure*}
\includegraphics[width=1.0\textwidth]{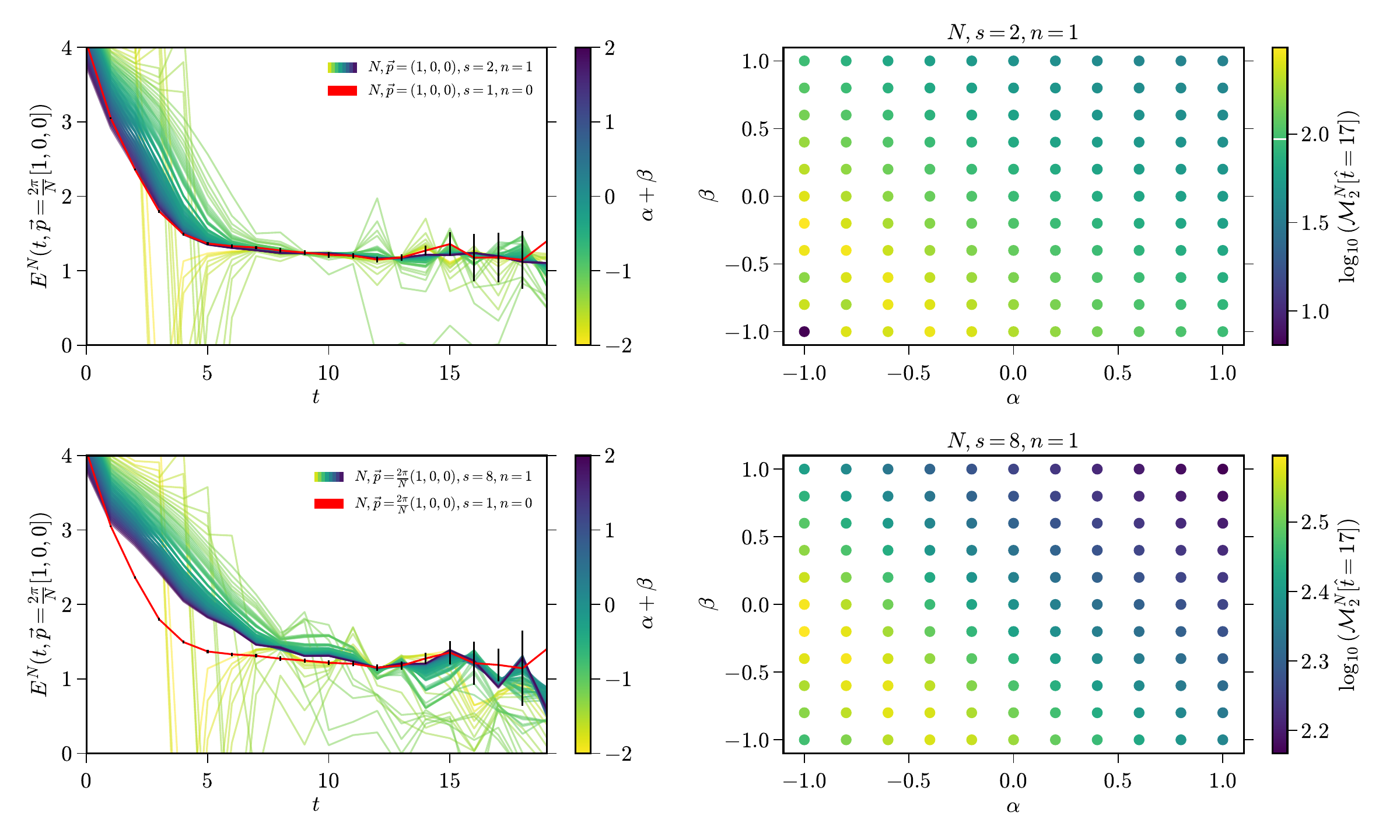}
    \caption{Examples of the typical effects observed in the two-point correlation functions as a function of the couplings in Eq. (\ref{eq:propagatorSparsening}) with 1 step of blocking applied to the sink. In this specific example, we show the proton effective-energy function at momentum $\vec{p}=\frac{2\pi}{N}[1,0,0]$. \textbf{Upper left:} Effective-energy function of the proton on a $12^3$ grid (decimation factor $s=2$) for $\alpha,\beta\in[-1,1]$. The color indicates the value of $\alpha+\beta$. \textbf{Lower left:} The same, but for a $3^3$ grid (decimation factor $s=8$). \textbf{Upper right: } Metric $\mathcal{M}_2$ (Eq. \eqref{eq:metrictwo}) as a function of $\alpha,\beta$ on a $12^3$ grid (decimation factor $s=2$). $\mathcal{M}_2$ computed for the unsparsened two-point correlation function is shown as a white line on the color-bar. \textbf{Lower right:} $\mathcal{M}_2$ for a $3^3$ grid (decimation factor $s=8$).}
    \label{fig:sampleTwoPoint}
\end{figure*}

\begin{figure*}
\includegraphics[width=1.0\textwidth]{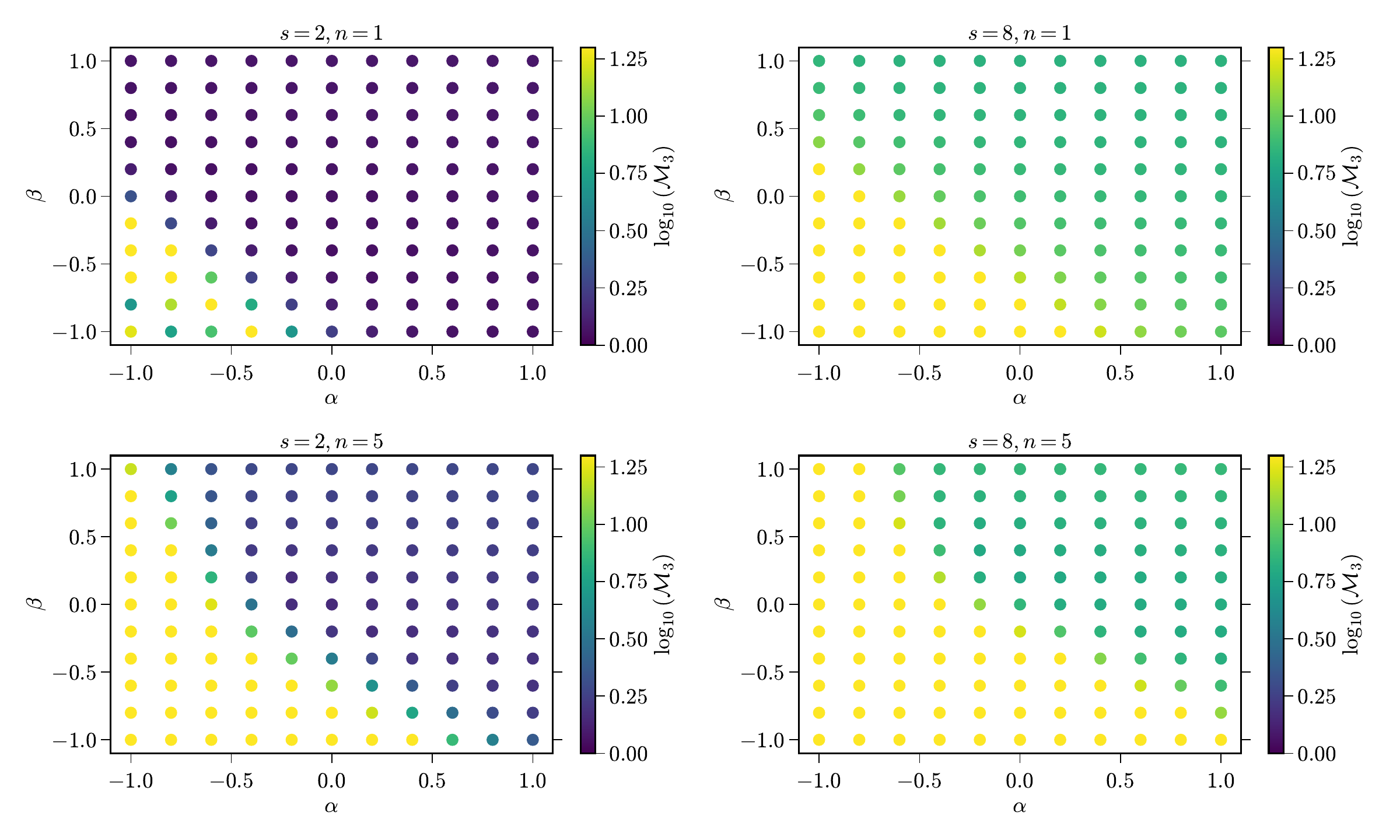}
    \caption{$\mathcal{M}_3$ (Eq. \eqref{eq:M3}) as a function of various couplings, blocking steps, and decimation levels. The couplings, blocking steps, and grids of the upper two panels correspond to Fig. \ref{fig:sampleTwoPoint}, while the lower two panels correspond to Fig. \ref{fig:sampleTwoPointMultipleSteps}. \textbf{Upper left:} The $\mathcal{M}_3$ metric as a function of $\alpha$ and $\beta$ for 1 blocking step and a $12^3$ grid. The color is saturated at 1.3, so any metrics with $\log_{10}(\mathcal{M}_3)\geq 1.3$ appear as the same color. \textbf{Upper right:} The same but for decimation to a $3^3$ grid. \textbf{Lower two panels:} The same but for 5 blocking steps.}
    \label{fig:variationStd}
\end{figure*}

A sample of the results with one step of blocking is shown in Fig. \ref{fig:sampleTwoPoint}. We present results from the proton, but these results are largely representative of those for the other hadrons (the zero-momentum pion correlation function is different in some respects because of constant ratio of signal-to-noise, however similar trends in $\mathcal{M}_{1,2,3}$ as a function of $\alpha$ and $\beta$ are observed). The average measurement errors, as quantified by $\mathcal{M}_3$ are shown as a function of $\alpha$ and $\beta$ in Fig.~\ref{fig:variationStd}.

A few trends are apparent in these results. Broadly speaking, we recover the previous results of Ref. \cite{Detmold:2019fbk}, where sparsening reproduces the ground-state energy robustly, but with more excited-state effects than the unsparsened correlation function because of the incomplete momentum projection. Since in most cases we are only sparsening the sink, the effective-energy function will not necessarily monotonically decrease in time as the correlation function is not convex. Indeed, we observe that for couplings approximately along the line $\alpha+\beta=-1$, there is non-monotonic behavior. 

\begin{figure*}
\includegraphics[width=1.0\textwidth]{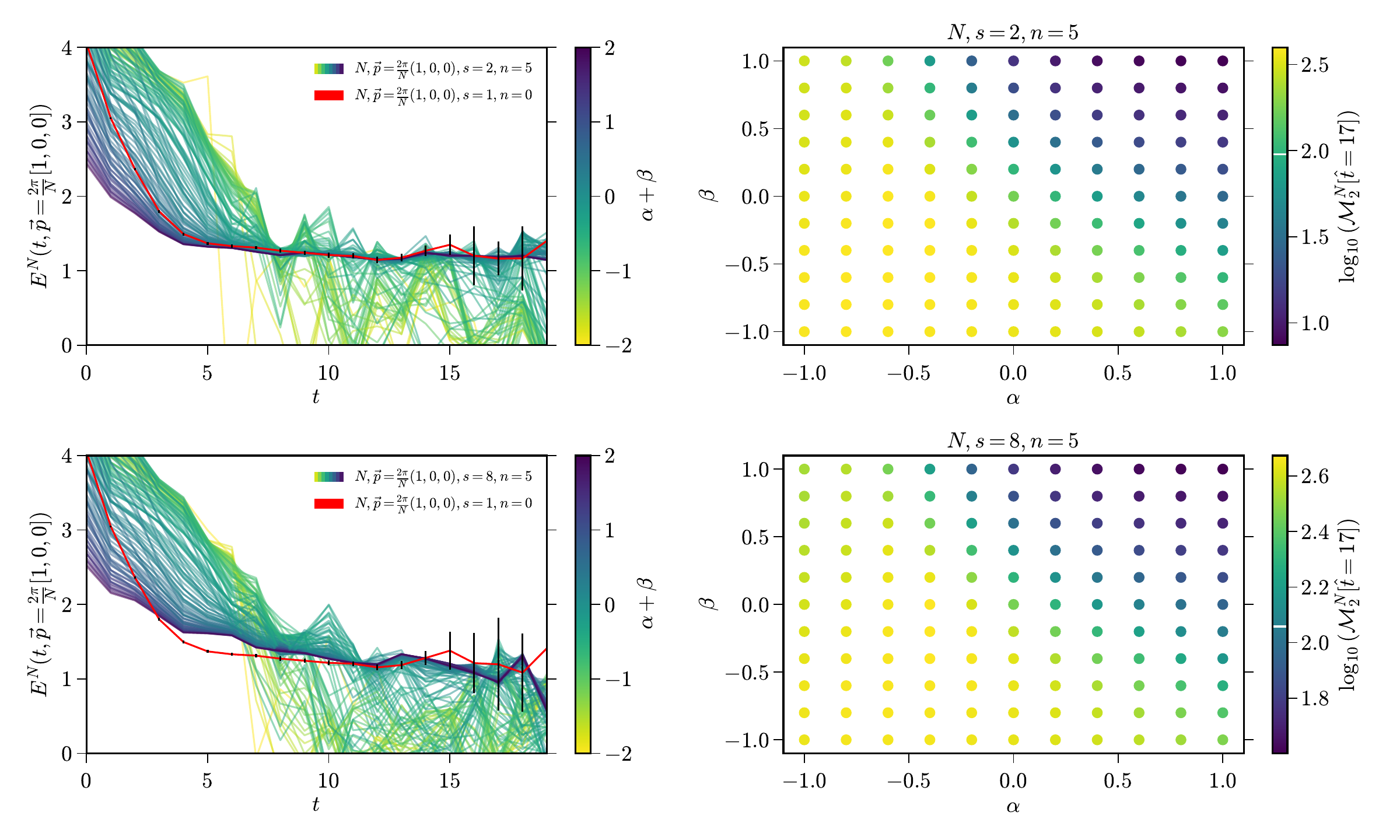}
    \caption{Examples of the typical effects observed in the two-point correlation functions as a function of the couplings in Eq. (\ref{eq:propagatorSparsening}) with 5 steps of blocking according to Eq. (\ref{eq:multipleBlockingSteps}) applied to the sink. In this specific example, we show the proton effective-energy function at momentum $\vec{p}=\frac{2\pi}{N}[1,0,0]$ analogously to Fig. \ref{fig:sampleTwoPoint}. \textbf{Upper left:} Effective-energy function of the proton on a $12^3$ grid (decimation factor $s=2$) for all $\alpha,\beta$. The color represents the value of $\alpha+\beta$. \textbf{Lower left:} The same, but for a $3^3$ grid (decimation factor $s=8$). \textbf{Upper right: } A plot of metric $\mathcal{M}_2$ (Eq. \eqref{eq:metrictwo}) as a function of $\alpha,\beta$ on a $12^3$ grid (decimation factor $s=2$). $\mathcal{M}_2$ computed for the unsparsened two-point correlation function is shown as a white line on the color-bar. \textbf{Lower right:} $\mathcal{M}_2$ for a $3^3$ grid (decimation factor $s=8$).}
    \label{fig:sampleTwoPointMultipleSteps}
\end{figure*}

We also consider multiple applications of blocking as in Eq. (\ref{eq:multipleBlockingSteps}), showing example results in Fig. \ref{fig:sampleTwoPointMultipleSteps}. For a small decimation of factor of $s=2$, higher-energy states are more suppressed with increasing applications of the blocking procedure, as can be seen in the upper left panel of Fig. \ref{fig:sampleTwoPointMultipleSteps}. For larger decimation factors such as $s=8$, the effective-energy functions for large coupling values are smaller at small times but then require a larger time to plateau than the unsparsened correlation function.
\begin{figure*}
\includegraphics[width=1.0\textwidth]{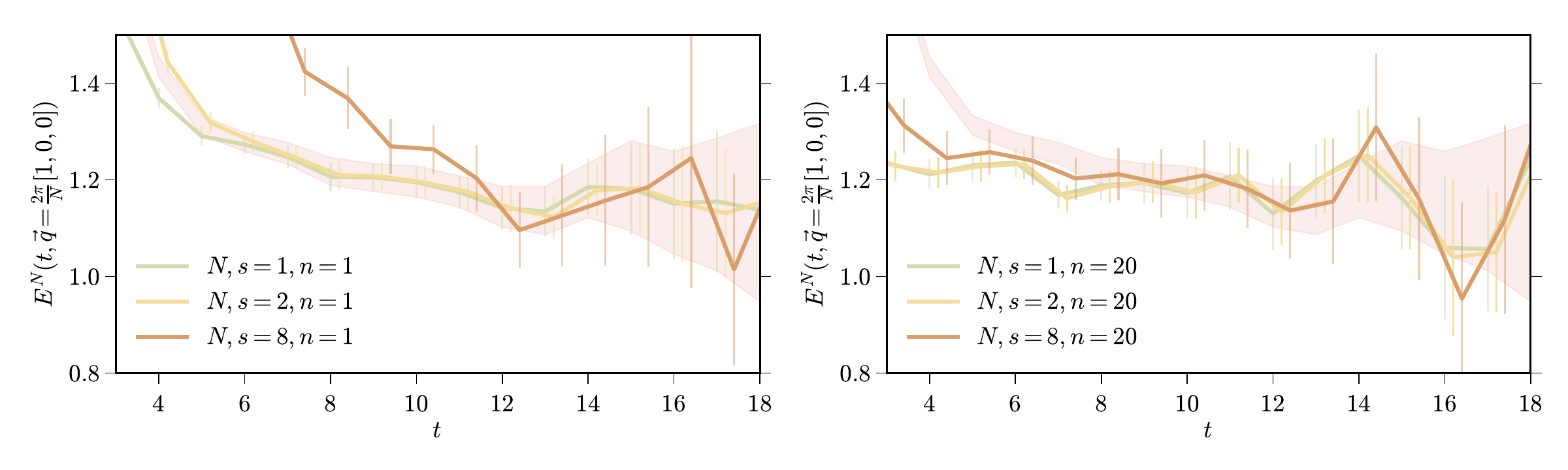}
    \caption{Examples of the typical effects observed in the two-point correlation functions as a function of the time extent. We show the proton effective-energy function at momentum $\vec{p}=\frac{2\pi}{N}[1,0,0]$ and use $\alpha=\beta=1$. For the case of $n=1$, we only block the sink but for $n=20$, we block both the source and sink. \textbf{Left:} Effective-energy function of the proton with one step of blocking $(n=1)$ for different decimation factors $(s=1,s=2,s=8)$. \textbf{Left:} Effective-energy function of the proton with 20 steps of blocking $(n=20)$ for different decimation factors $(s=1,s=2,s=8)$.
    In both panels, the pink shaded-region corresponds to the unsparsened correlation function and its uncertainties.}
    \label{fig:blockingProtonExamples}
\end{figure*}
\begin{figure*}
\includegraphics[width=1.0\textwidth]{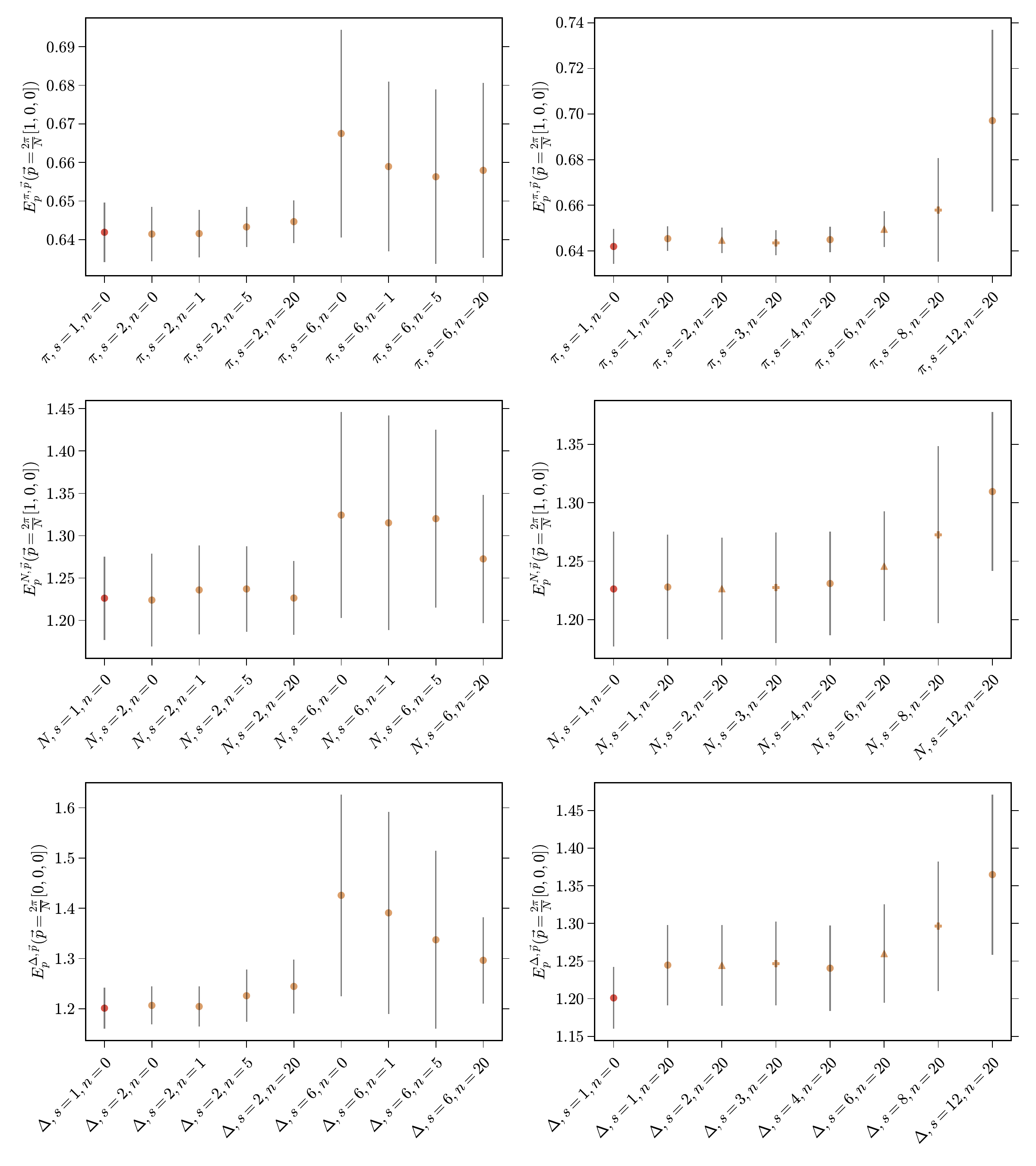}
    \caption{The best-fit values of the pion, proton, and $\Delta$ baryon effective energy $E_p^{h,\vec{p}}$. We use $\alpha=\beta=1$ throughout. In the case of $n=1$ and 5, we block only the sink but for $n=20$, we block both the source and sink. \textbf{Left:} $E_p^{h,\vec{p}}$ with varying steps of blocking $(n=1,n=5,n=20)$ and decimation factors $(s=1,s=2,s=8)$. \textbf{Right:} Effective energies of the pion, proton, and $\Delta$ baryon with 20 steps of blocking $(n=20)$ for different decimation factors $(s=1,s=2,s=8)$.
    }
    \label{fig:plateauProtonEnergy}
\end{figure*}
The fact that correlation functions built from sparsened propagators for small $s$ and $\alpha,\beta\neq 0$ recover the ground-state energy at earlier times than a naive point source is not a surprising result and is similar to previous results found for smearing. The results are reflective of the blocking step of our analysis, and not the sparsening process which also includes decimation.

Because the large number of smearing steps efficiently eliminates higher-energy contributions, for relatively small decimation factors, the effective energy is almost exactly the same as the undecimated -- but still blocked -- effective energy. This effect can be seen in Fig. \ref{fig:blockingProtonExamples} which summarizes the results of applying 1 versus 20 blocking steps on the proton effective energy at momentum $\vec{p}=\frac{2\pi}{N}[1,0,0]$. Applying 20 steps of blocking decreases excited state contamination, in-line with previous results for operator smearing. However, we also see improved agreement between undecimated and decimated results for 20 blocking steps compared to 1 blocking step, suggesting that one can generally utilize larger decimation factors for 20 blocking steps versus 1 blocking step. This result is in line with the expectation that sparsenings with more blocking steps retain more long-distance information after decimation.

In Fig. \ref{fig:plateauProtonEnergy}, we show the best-fit energy of the pion, proton, and $\Delta$ two-point correlation function at momentum $\vec{p}=\frac{2\pi}{N}[1,0,0]$ for the case of the pion and proton, and $\vec{p}=[0,0,0]$ for the case of the delta baryon according to the procedure of section \ref{sec:methodsOfSuccess} for varying blocking and decimation factors. Comparing the measurement uncertainty on the best-fit energy as a function of blocking and decimation step shows the competing effects of both increasing statistical uncertainty and decreasing excited states with increasing blocking steps. On the left-hand side of this figure, we see that in the case of the proton at momentum $\vec{p}=\frac{2\pi}{N}[1,0,0]$, the measurement uncertainty in the best-fit energy decreases as the number of blocking steps is increased. 
On the right-hand side of this figure, we show the best-fit energy for all possible decimations at a fixed blocking $(n=20)$. We observe that as the number of blocking steps is increased, the best-fit energy becomes increasingly biased away from its undecimated value, although it remains consistent within one standard deviation of the undecimated result. 

\begin{table}\begin{ruledtabular}\begin{tabular}{ccc}
  $s$ & $n$   & $\overline\sigma$ \\ \hline
   1 & 0  & 0.022 \\
   2 & 0 & 0.022 \\
   2 & 1 & 0.021 \\
   2 & 5 & 0.041 \\
   2 & 20 & 0.064
\end{tabular}
\end{ruledtabular}
\caption{The mean standard deviation of the proton effective-energy function for $\vec{p}=\frac{2\pi}{N}[1,0,0]$ between $t=1$ and 10, $\overline{\sigma}$,  computed using the bootstrap procedure for various levels of blocking and decimation for sparsening both the source and sink with $\alpha=\beta=1$.
\label{tab:errors}}
\end{table}
A trade-off with additional blocking steps is that they tend to increase the statistical uncertainty. In Table \ref{tab:errors}, we show the average uncertainty in the proton effective-energy function for $\vec{p}=\frac{2\pi}{N}[1,0,0]$ for $\alpha=\beta=1$ as a function of blocking step.  Additionally, $\mathcal{M}_3$ (Eq.~\eqref{eq:M3}), a measure of the average error present for given couplings and blocking steps, is shown in Fig.~\ref{fig:variationStd} for all $\alpha,\beta$. In both cases, we see that as the number of blocking steps is increased, the statistical uncertainty typically  increases as well.

\section{Three-point correlation functions}
\label{sec:three}

Having investigated more general forms of sparsening in two-point correlation functions, we now turn to three-point correlation functions used to determine hadronic matrix elements, focusing in particular on the vector current and the corresponding electromagnetic form factors.

Three-point correlation functions involving operator insertions can in principle be sparsened at all three coordinate-locations: the operator insertion, and both the source or sink hadronic interpolating operators. 
Sparsening at the operator changes the matrix element that is being computed in non-perturbative ways and will require additional renormalization. In most scenarios, it is also not clear how sparsening at the operator insertion leads to significant cost savings (see Ref.~\cite{Davoudi:2024ukx} for a related discussion). 
Consequently, in this study we construct sequential propagators through the operator insertion and then compute three-point correlation functions sparsened at the source and sink that involve the sequential propagator. 

\subsection{Details of lattice calculations}
\label{sec:threePointSetup}

We compute three-point correlation functions of the pion and proton with vector-current operator insertions of the form
\begin{align}
\label{eq:Amu}
    C_3^h(t,\tau, \vec{p},\vec{q})&\equiv\langle\mathcal{O}_h^\dagger(\vec{p},t)V_4(\vec{q},\tau)\mathcal{O}_h(\vec{x}=0,0)\rangle,
\end{align}
where $V_\mu(\vec{q},\tau)=\sum_{\vec{x}}e^{-i\vec{q}\cdot\vec{x}}\overline{q}(\vec{x},\tau)\gamma_\mu q(\vec{x},\tau)$ corresponds to the vector current in momentum space.
We write $\vec{x}=0$ to emphasize that the source operator is not momentum projected. In the large-time limit, it is straightforward to show that
\begin{multline}
\label{eq:wellSepLimitThreePoint}
    C_3^h(t,\tau,\vec{p},\vec{q})\stackrel{0\ll \tau \ll t \ll T/2}{\longrightarrow} e^{-E^{h,\vec{p}}(t-\tau)}e^{-E^{h,\vec{p}+\vec{q},}\tau}\\
    \times\langle \Omega|\mathcal{O}_h^\dagger(\vec{p},t)|h\rangle\langle h|V_4(\vec{q},\tau)|h'\rangle\langle h'|\mathcal{O}_h(\vec{p}+\vec{q},0)|\Omega\rangle,
\end{multline}
where $|h\rangle$ and $|h'\rangle$ are the lowest energy states with the appropriate quantum numbers and momenta, and $|\Omega\rangle$ is the vacuum state. Lorentz invariance implies $\langle h|V_\mu(\vec{q},\tau)|h'\rangle$ can be written in terms of known kinematic quantities and form factors that are functions of only $q^2$. For simplicity, we consider only the coupling to the $u$ quark,  $V_\mu(\vec{q},\tau)= \sum_{\vec{x}}e^{-i\vec{q}\cdot\vec{x}}\overline{u}(\vec{x},\tau)\gamma_\mu u(\vec{x},\tau)$, ignore disconnected insertions, and set $\vec{p}=0$.

Sparsening modifies the normalizations of two- and three-point correlation functions differently and care must be taken to account for this.
The essential point is that we recover the matrix element of interest for sufficiently separated $t,\tau,T$ in the case that any component of  $\vec{q}$ is not equal to $ \frac{\pi}{s}$ from the ratio:
\begin{align}
\label{eq:formFactorDef}
    &R^h(t,\tau,\vec{q})\equiv\frac{C_3^h(t,\tau,\vec{0},\vec{q})}{C^h(t,\vec{0})}\sqrt{\frac{C^h(t,\vec{0})C^h(\tau,\vec{0})C^h(t-\tau,\vec{q})}{C^h(t,\vec{q})C^h(\tau,\vec{q})C^h(t-\tau,\vec{0})}},
\end{align}
where in particular
\begin{align}
    R^h(t,\tau,\vec{q})
    \stackrel{{0\ll \tau \ll t\ll T/2}}{\longrightarrow}
    \langle h|V_4(\vec{q},\tau)|h'\rangle.
\end{align}
This quantity allows combinations of the electromagnetic form-factors of hadron $h$ to be determined, but it involves both the two- and three-point correlation functions. We focus on this ratio in our numerical investigations. One could alternatively compare the ratio of the sparsened three-point correlation function divided by the original, unsparsened two-point correlation function, but differences in normalization would be seen due to the blocking and decimation steps. This normalization cannot be simply corrected for (e.g., multiplying by the number of points lost due to decimation) because of the effects of the relative values of the covariantly-transported neighboring points controlled by $\alpha,\beta$.
Indeed, Eq.~\eqref{eq:formFactorDef} must be modified if any component of the momentum vector is $\vec{q}$ is $\frac{\pi}{s}$, which according to the analysis in Sec.~\ref{sec:idealLimit} is the maximum momentum that can be accessed using sparsening. At this momentum, the two-point correlation function asymptotes as
\begin{align}
    C^h\left(t,\vec{q} =\left(\frac{\pi}{s},q_\perp\right)\right)&
    \stackrel{t\to\infty}{\to}
    c e^{-E^h(\frac{\pi}{s},q_\perp)t}+c e^{-E^h({2\pi}-\frac{\pi}{s},q_\perp)t}\notag\\
    &=2c e^{-E^h(\frac{\pi}{s},q_\perp)t},
\end{align}
where $c$ is an overlap factor. This analysis is similar to that of Sec.~\ref{sec:idealLimit} for momenta greater than $\frac{\pi}{s}$, where different momentum-modes dominate the result due to the periodic dispersion-relation. Here, there are $2m$ momenta values that contribute equally, where $m$ is the number of Cartesian momenta components equal to $\frac{\pi}{s}$. In our numerical studies, $m=1$ at most and therefore, the  ratio in Eq.~\eqref{eq:formFactorDef} is modified as
\begin{align}
\label{eq:modifiedFormFactorDef}
    R^h(t,\tau,\vec{q}=(\frac{\pi}{s},q_\perp))&\equiv\sqrt{2}\frac{C_3^h(t,\tau,\vec{0},\vec{q})}{C^h(t,0)}\notag\\
    &\times\sqrt{\frac{C^h(t,\vec{0})C^h(\tau,\vec{0})C^h(t-\tau,\vec{q})}{C^h(t,\vec{q})C^h(\tau,\vec{q})C^h(t-\tau,\vec{0})}}.
\end{align}

As in the analysis of two-point correlation functions, we find the best-fit plateau to $R^h(t,\tau,\vec{q})$ at fixed $\tau$ and $\vec{q}$ according to the procedure of Sec.~\ref{sec:methodsOfSuccess}. We denote this quantity $R_p^h(\tau,\vec{q})$. Since the form factor is extracted from the $0\ll\tau\ll t\ll T/2$ limit, we also study the effects of sparsening as $\tau$ is varied.

\subsection{Results}
\label{sec:threePointResults}

\begin{figure*}
\includegraphics[width=1.0\textwidth]{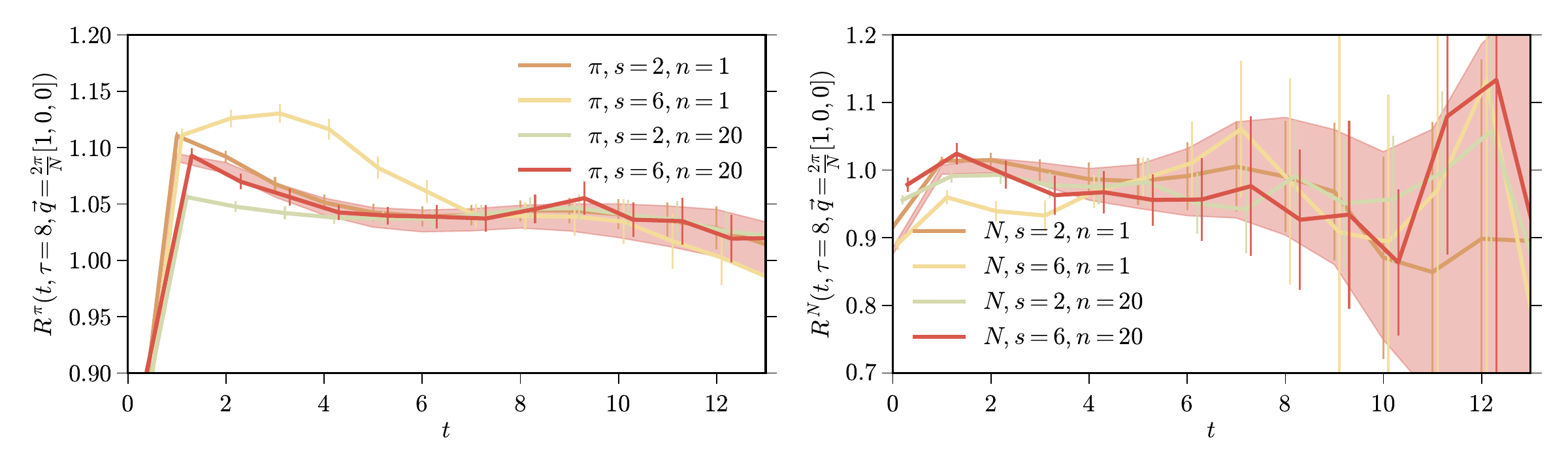}
    \caption{Examples of the typical effects observed in the improved ratio of the three- and two-point correlation functions, $R^h(t,\tau,\vec{q}=\frac{2\pi}{N}[1,0,0])$. Throughout, we set $\alpha=\beta=1$ and $\tau=8$. \textbf{Left:} $R^\pi$ as a function of sink time $t$ for four choices of blocking and decimation factors,  $n\in\{1,20\}$ and $s\in\{2,6\}$. For $n=1$ we only block the sink but for $n=20$ we block both the source and sink. The pink shaded-region shows the mean and uncertainty of the unblocked, undecimated form-factor. \textbf{Right:} The same parameter variations are shown for $R^N$. }
    \label{fig:formFactorExamples}
\end{figure*}

\begin{figure*}
\includegraphics[width=1.0\textwidth]{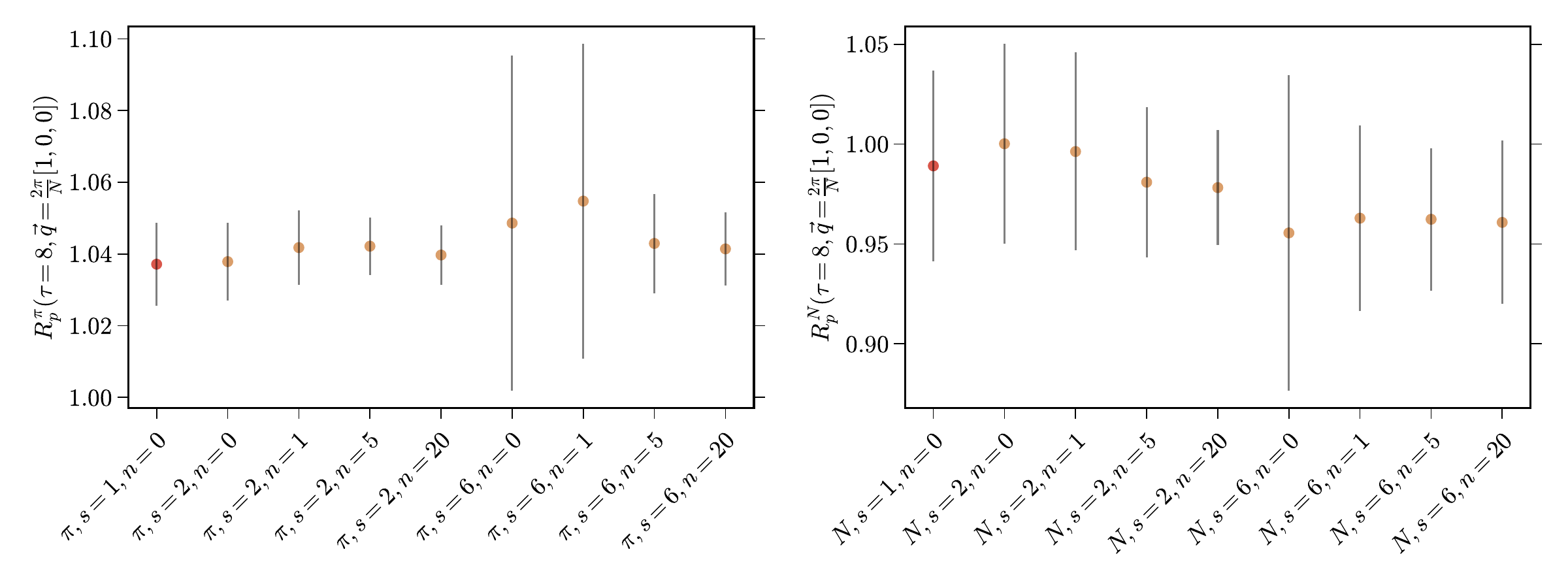}
    \caption{Best-fit plateau values $R_p^h$ (Eq.~\eqref{eq:formFactorDef}) determined using the procedures of Sec.~\ref{sec:methodsOfSuccess}. \textbf{Left}: The best-fit plateau of $R_p^\pi$ at momentum $\vec{q}=\frac{2\pi}{N}[1,0,0]$ and varying decimation factors ($s=2,s=6$) and blocking factors ($n=0,n=1,n=5,n=20$). The unblocked, undecimated result is shown in red. \textbf{Right}: The best fit plateau values $R_p^N$ for the same parameter variations.}
    \label{fig:formFactorPlateaus}
\end{figure*}

The metrics $\mathcal{M}_1$, $\mathcal{M}_2$, and $\mathcal{M}_3$ constructed from $R^h(t,\tau,\vec{q})$ as a function of $\alpha,\beta$ and number of blocking steps and decimation factors largely follow the patterns found from the two-point correlation functions. For example, the couplings that give the best agreement with the ground state tend to be near $\alpha=\beta=1$. For one step of blocking, the couplings that introduce the most excited states and deviate the most from a positive-definite function tend to be along the line $\alpha+\beta=-1$.

In contrast to the two-point correlation functions, there are differences in the efficacy of the sparsening procedure between the three-point correlation functions for the pion and the proton. In Fig.~\ref{fig:formFactorExamples}, we show example blocking and decimation steps for the case of the pion and proton. In Fig.~\ref{fig:formFactorPlateaus}, we show the best-fit values to $R_p^\pi$ and $R_p^N$ determined according to the procedure of Sec. \ref{sec:methodsOfSuccess}. Comparing these results, we observe that the $R_p^h$ value is recovered to some degree in both the pion and proton when blocking is applied, but the uncertainty on the plateau for the pion decreases significantly with increasing blocking. In Fig.~\ref{fig:formFactorExamples}, we see that this is because for $R^N$, statistical uncertainties dominate over excited-state effects.\footnote{Note that in this study, the two-point correlation functions in Eq.~\eqref{eq:formFactorDef} have all plateaued to their ground state at the times that are studied. If not, there could be cancellations in the decay of two- and three-point correlation functions, leading to a spurious plateau in $R^h$.}
In the case of $R^N$ with a large number of blocking steps ($n=20$), the effects of  excited states are subdominant to the statistical uncertainties. Therefore, blocking is less effective for $R^N$ than for $R^\pi$.

\begin{figure*}
\includegraphics[width=1.0\textwidth]{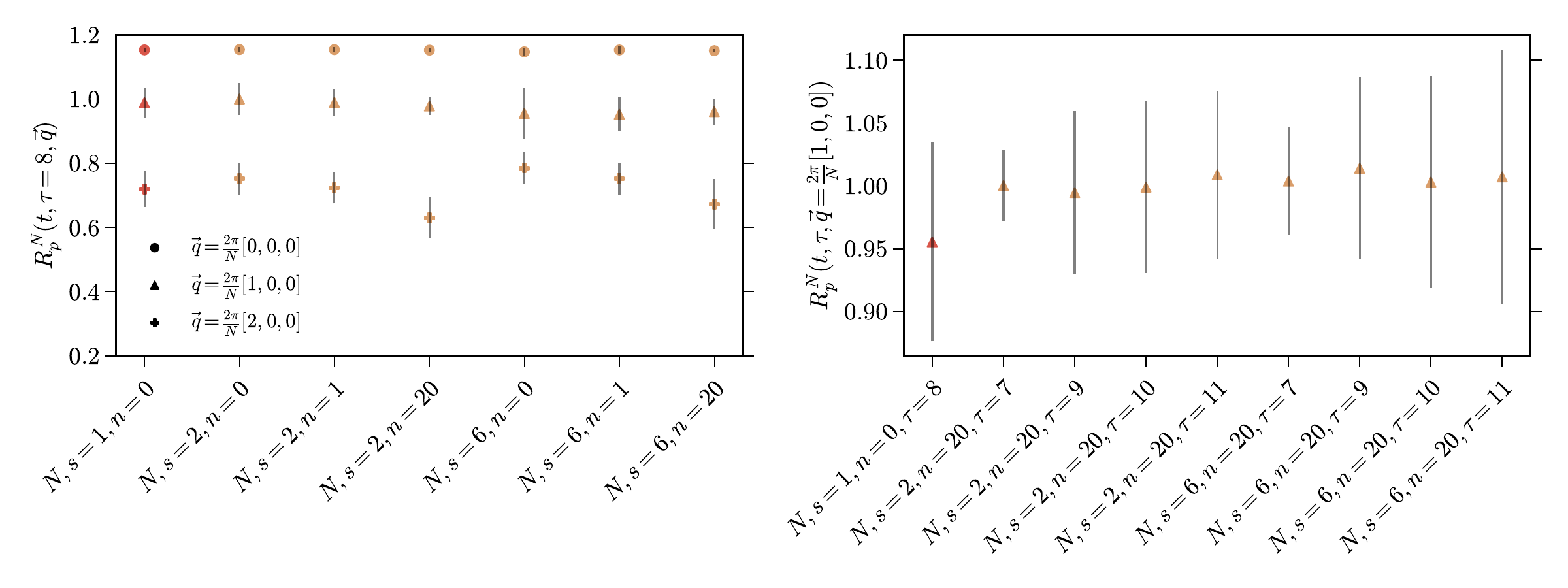}
    \caption{Best-fit plateau values $R_p^N$ (Eq.~\eqref{eq:formFactorDef} and, for the last three results of the left hand panel, Eq.~\eqref{eq:modifiedFormFactorDef}) according to the procedure of Sec.~\ref{sec:methodsOfSuccess} for various momenta and operator insertion times $\tau$. \textbf{Left}: The best-fit plateau of $R_p^N$ for varying momenta ($\vec{q}\in{\frac{2\pi}{N}}\{[0,0,0],[1,0,0],[2,0,0]\}$) and varying decimation factors ($s=2,s=6$) and blocking factors ($n=1,n=20$). The unblocked, undecimated result is shown in red. \textbf{Right}: $R^N_p$ for varying $\tau \in \{7,9,10,11\}$ and decimation factors ($s=2,s=6$) at fixed $n=20$ and $\vec{q}={\frac{2\pi}{N}[1,0,0]}$.}
    \label{fig:formFactorDifferentMomenta}
\end{figure*}

In Fig.~\ref{fig:formFactorDifferentMomenta}, we show the best-fit values for $R^N_p$ as a function of varying momenta and operator insertion times $\tau$ for a range of blocking and decimation factors. We find that sparsening robustly recovers the best-fit values as a function of momentum. Additionally, we observe that as we vary $\tau$, the best-fit plateau of $R^N_p$ remains relatively constant, indicating we are studying the region of $R^N(t,\tau,\vec{q})$ where $t$ and $\tau$ are sufficiently well-separated such that the ground states dominate before and after the operator insertion.

\section{Discussion}
\label{sec:analysis}

Some overall observations on the effects of the more general sparsening procedure that was introduced here can be made based on the preceding studies of two- and three-point correlation functions.

\begin{enumerate}
\item Figures \ref{fig:sampleTwoPoint} and \ref{fig:sampleTwoPointMultipleSteps} demonstrate that across varying number of blocking steps, the ground state of the two-point correlation function is recovered at increasingly earlier times for larger $\alpha,\beta$. These figures also demonstrate that sparsening with $\alpha+\beta<0$ generally does not recover the ground state effectively, as measured through the metric $\mathcal{M}_2$.

\item  For $\alpha=\beta=1$, Figs. \ref{fig:blockingProtonExamples} and \ref{fig:formFactorPlateaus} demonstrate that increasing the number of blocking steps recovers the ground state of the two-point correlation function and the optimized ratio of three-point to two-point correlation functions at earlier times across different amounts of decimation. In both examples, as time increases, the magnitude of the statistical uncertainty also increases.

\item  One way to measure the trade-offs of these two competing effects is to determine the ground state of the respective correlation functions (Sec. \ref{sec:methodsOfSuccess}), and then compare the statistical uncertainties of this extracted ground state across different sparsening methods. Figures \ref{fig:plateauProtonEnergy}, \ref{fig:formFactorPlateaus}, and \ref{fig:formFactorDifferentMomenta} demonstrate that for large amounts of decimation ($s=6$), increasing the number of blocking steps decreases the statistical uncertainty in extracted ground states.

\item Finally, across all quantities that we computed, the effects of increasing $\alpha$ and $\beta$ are very similar, as measured through the $\mathcal{M}_2$ metric. More precisely, the results of sparsening for $(\alpha,\beta)=(A,B)$ and $(\alpha,\beta)=(B,A)$  produce sparsened correlation functions with absolute differences of $\log_{10}(\mathcal{M}_2)$ less than 0.1 in almost all cases. Moreover, for one step of blocking, the right hand side of Fig. \ref{fig:sampleTwoPoint} demonstrates that the least effective sparsening (as measured through the metric $\mathcal{M}_2$) is for $\alpha,\beta$ near the line $\alpha+\beta=-1$. As discussed in Appendix \ref{sec:free}, this is similar to the results obtained for a free propagator. At small separations where QCD becomes asymptotically free, sparsening parameters around $\alpha+\beta=-1$ lead to nearly total cancellation of the nearest-neighbor and next-to-nearest-neighbor sites against the central site.
\end{enumerate}

\section{Conclusion}

In this study, we have tested a more general form of propagator sparsening, focusing on various combinations of sequential blocking (sequentially covariantly-averaging propagators) and decimation (discarding a certain set of sites to reduce the spatial lattice size). The results of this study suggest that sparsening with sequential blocking is effective at preserving long-distance correlations in the sparsened propagator and therefore at preserving low-energy components of hadronic correlation functions at high fidelity. Consequently, larger decimations can be used, reducing storage requirements and subsequent computational costs of Wick contractions.
In deciding on an optimal choice of sparsening, the competing effects of suppression of excited states and additional noise must be balanced against these computational improvements.

\acknowledgments

We are grateful to Christoph Lehner for helpful discussions.
Computations in this work were carried out using the Chroma~\cite{Edwards:2004sx} and gpt/grid~\cite{gpt,Boyle:2016lbp} software libraries. 

This research used resources of the Oak Ridge Leadership Computing Facility at the Oak Ridge National Laboratory, which is supported by the Office of Science of the U.S. Department of Energy under Contract number DE-AC05-00OR22725
and the resources of the National Energy Research Scientific Computing Center (NERSC), a Department of Energy Office of Science User Facility using NERSC award NP-ERCAPm747. 
The authors also acknowledge the MIT SuperCloud \cite{reuther2018interactive} and Lincoln Laboratory Supercomputing Center for providing HPC resources that have contributed to the research results reported within this paper.

This work was supported in part by the U.S. Department of Energy, Office of Science under grant Contract Number DE-SC0011090, by the SciDAC5 award DE-SC0023116, and by the MIT Undergraduate Research Opportunities Program (UROP).

\appendix

\section{Free propagator}
\label{sec:free}
In this appendix, we perform the same analysis for a free lattice-fermion action as was performed for QCD. To closely mirror our previous analysis, we use the same lattice geometry and use the Wilson-clover fermion action. For this toy model, instead of antiperiodic boundary conditions in the temporal direction, we use periodic boundary conditions so the gauge field can be set to the identity everywhere.

We perform our analysis directly for the free-fermion propagator since any other correlation functions built from fermion field operators are simply determined from this by Wick's theorem. We use a bare quark mass of $0.3$ and vary $\alpha,\beta\in [-1,2]$.
\begin{figure*}
\includegraphics[width=1.0\textwidth]{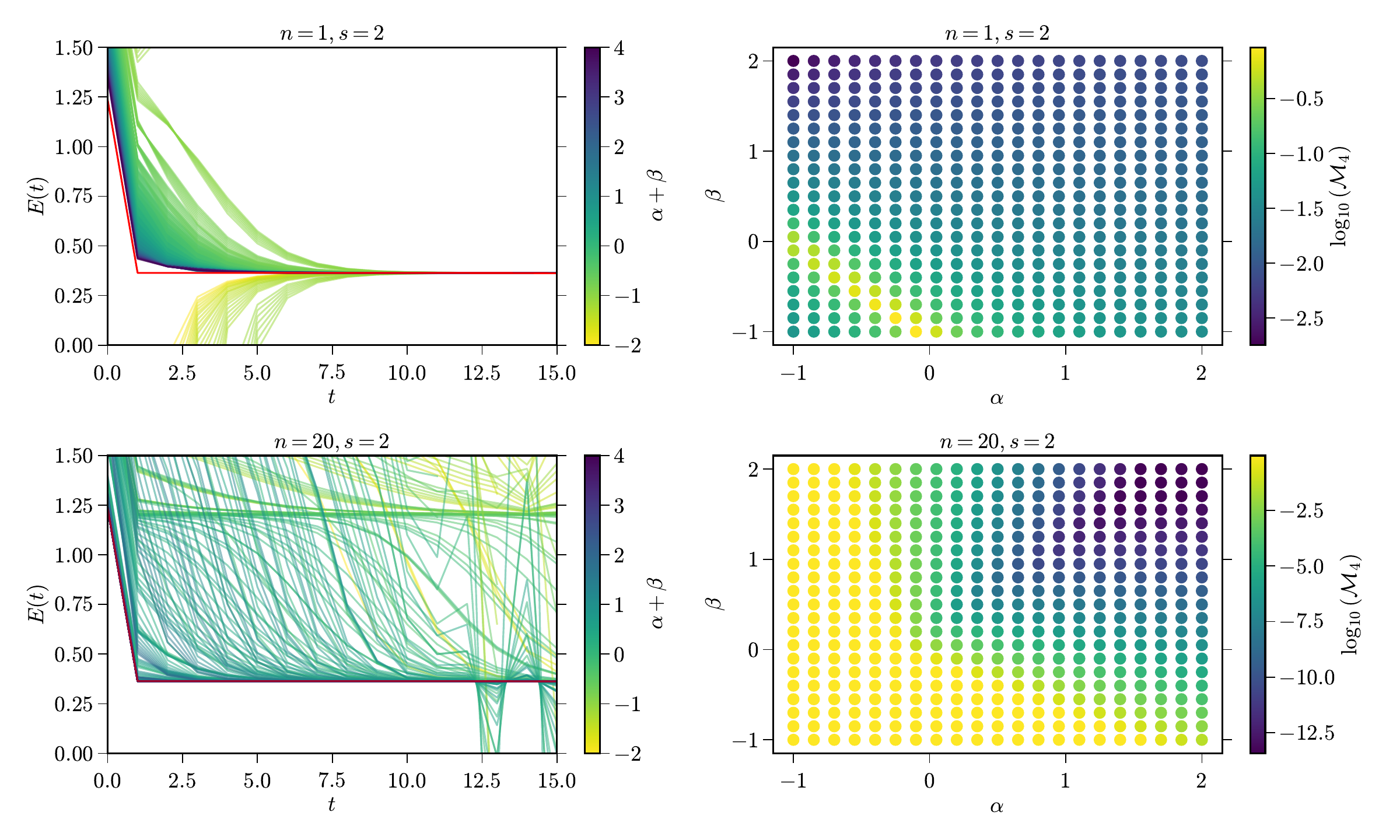}
    \caption{Defining features of results computed from a free propagator at momentum $\vec{p}=\frac{2\pi}{N}[1,0,0]$. \textbf{Upper left:} The effective energy of the free propagator with 1 blocking step and $s=1$ decimation to a $12^3$ spatial grid for varying $\alpha,\beta$. We compare to the original, unsparsened result in red. \textbf{Lower left:} The same but with 20 blocking steps. \textbf{Upper right:} The energy difference at time $t=5$ for the sparsened propagator of the upper left compared to the original unsparsened propagator. \textbf{Lower right:} The same but for 20 blocking steps, corresponding to the lower left.}
    \label{fig:freeExamples}
\end{figure*}
To quantity our results in the absence of noise, we use the metric
\begin{align}
    \mathcal{M}_4(t)&=|E_{us}(t)-E_s(t)|.
\end{align}

The results of this experiment are shown in Fig. \ref{fig:freeExamples} for $\vec{p}=\frac{2\pi}{N}[1,0,0]$. Comparing the upper-right of this figure to the upper-right and lower-right of Fig. \ref{fig:sampleTwoPoint}, we see qualitative similarities. In particular, both have slope $-1$ in how the magnitude varies with coupling, and the sparsened results show large disagreement with the unsparsened results around $\alpha+\beta= -1$. This similarity suggests that in the QCD case, the nearest- and next-to-nearest-neighbor terms are largely probing the asymptotically-free regime of the propagator.

Similarities are also seen when a large number of blocking steps are applied. Comparing the lower-right of Fig.~\ref{fig:freeExamples} and the right-hand side of Fig.~\ref{fig:sampleTwoPointMultipleSteps} reveals qualitative similarities. For example, the sparsened free-propagator results reproduce the unsparsened result more and more accurately as the number of blocking steps is increased for positive couplings. This is similar to the example in Fig.~\ref{fig:sampleTwoPointMultipleSteps}, although in the case of QCD, increased blocking also results in increased statistical uncertainty. Moreover, this figure confirms the theoretical expectations in Eq.~\eqref{eq:secondStep}. Instead of decaying to the ground state energy of $0.262$, the left-hand side of the figure shows that some choices of $\alpha$ and $\beta$ result in $E(t,\vec{p})$ temporarily plateauing to $1.193$. For the unsparsened dispersion relation, this corresponds to $\vec{p}=\frac{2\pi}{N}(13,0,0)$, which is the second contribution in Eq.~\eqref{eq:secondStep}. For these choices of $\alpha$ and $\beta$, the coefficient for the lowest momentum contribution in Eq.~\eqref{eq:secondStep} is very small such that the second contribution dominates for larger times.

\bibliography{refs}

\end{document}